\documentclass[pdflatex,sn-mathphys]{sn-jnl}

\jyear{2021}%

\theoremstyle{thmstyleone}%
\theoremstyle{thmstyletwo}%

\theoremstyle{thmstylethree}%

\begin{document}

\title[EEG Opto-processor]{EEG Opto-processor: epileptic seizure detection using diffractive photonic computing units}



\author[1,3]{Tao Yan}
\author[2,4]{Maoqi Zhang}
\author[1,3,5]{Sen Wan}
\author[1,3]{Kaifeng Shang}
\author[2]{Haiou Zhang}
\author[4]{Xun Cao}
\author*[2,3,5]{Xing Lin}\email{lin-x@tsinghua.edu.cn}
\author*[1,3,5]{Qionghai Dai}\email{daiqh@tsinghua.edu.cn}

\affil[1]{Department of Automation, Tsinghua University, Beijing 100084, China}
\affil[2]{Department of Electronic Engineering, Tsinghua University, Beijing 100084, China}
\affil[3]{Institute for Brain and Cognitive Sciences, Tsinghua University, Beijing 100084, China}
\affil[4]{School of Electronic Science and Engineering, Nanjing University, Nanjing 210023, China}
\affil[5]{Beijing National Research Center for Information Science and Technology, Tsinghua University, Beijing 100084, China}


\abstract{Electroencephalography (EEG) analysis extracts critical information from brain signals, which has provided fundamental support for various applications, including brain-disease diagnosis and brain-computer interface. However, the real-time processing of large-scale EEG signals at high energy efficiency has placed great challenges for electronic processors on edge computing devices. Here, we propose the EEG opto-processor based on diffractive photonic computing units (DPUs) to effectively process the extracranial and intracranial EEG signals and perform epileptic seizure detection. The signals of EEG channels within a second-time window are optically encoded as inputs to the constructed diffractive neural networks for classification, which monitors the brain state to determine whether it's the symptom of an epileptic seizure or not. We developed both the free-space and integrated DPUs as edge computing systems and demonstrated their applications for real-time epileptic seizure detection with the benchmark datasets, i.e., the CHB-MIT extracranial EEG dataset and Epilepsy-iEEG-Multicenter intracranial EEG dataset, at high computing performance. Along with the channel selection mechanism, both the numerical evaluations and experimental results validated the sufficient high classification accuracies of the proposed opto-processors for supervising the clinical diagnosis. Our work opens up a new research direction of utilizing photonic computing techniques for processing large-scale EEG signals in promoting its broader applications.}




\maketitle

\section{Introduction}\label{sec1}

Electroencephalography (EEG) monitors the brain's neural activities by measuring the electrical fields with electrons, where the extracranial EEG places electrodes outside the skull and intracranial EEG, i.e., iEEG, implants the electrodes directly on the cerebral cortex. The complexity of EEG signals requires advanced signal processing and analysis methods for correct interpretation. Recently, deep learning \cite{lecun2015deep} has shown significant advantages in analyzing EEG signals by learning the feature representation and data abstraction to facilitate different applications, e.g., epilepsy, brain-computer interface, sleep, and cognitive monitoring \cite{craik2019deep,gao2021complex,schirrmeister2017deep}. The deep learning architectures based on artificial neural networks (ANNs) are generally implemented on electronic computing platforms using central processor units (CPUs), graphics processing units (GPUs), field-programmable gate arrays (FPGAs), and application-specific integrated circuits (ASICs) \cite{yao2020fully}. With the continuously increasing EEG signal channels and data scale, processing the EEG signals using deep learning on electronic computing platforms are power-intensive and time-consuming. Therefore, it's challenging to deploy the models on edge devices for portable and wearable applications due to power supply constraints and limited computing speed. Besides, electronic computing confronts unsustainable performance growth as electronic transistors approach their physical limits \cite{waldrop2016chips}.

Photonic computing has been considered a promising solution for future computing systems due to its significant advantages of high energy efficiency and light-speed parallel processing capabilities \cite{caulfield2010future}. Using photonic computing, research on photonic neural networks (PNNs) and photonic integrated circuits have flourished vigorously in recent years \cite{wetzstein2020inference, shastri2021photonics}, which enable ultra-fast and low-power artificial intelligence (AI) inference and provide critical support for high-performance computing in scenarios with strict limits on energy consumption. Various effective architectures of PNNs have been demonstrated, including spike neural networks \cite{feldmann2019all}, convolutional neural networks \cite{chang2018hybrid, miscuglio2020massively}, reinforcement learning \cite{bueno2018reinforcement}, and reservoir computing \cite{antonik2019human}. The linear weighted interconnection of PNNs could be implemented via a mesh of Mach–Zehnder interferometers (MZIs) \cite{shen2017deep}, diffractive surfaces \cite{lin2018all, yan2019fourier, rahman2021ensemble, kulce2021all, veli2021terahertz, zhou2021large, doi:10.1126/sciadv.abn7630}, and tunable attenuators connected with waveguides \cite{tait2017neuromorphic,feldmann2019all}, etc. The nonlinear activation function of PNNs could be realized by laser-cooled atoms with electro-magnetically induced transparency \cite{zuo2019all}, cavity-loaded interferometers \cite{jha2020reconfigurable}, and photodetector-driven MZIs \cite{williamson2019reprogrammable}, etc. Currently, photonic accelerators enable massive parallel computing at speeds of trillions of multiply-accumulate operations per second \cite{feldmann2021parallel, xu202111}. But most existing PNNs only support simple neural network architectures with limited numbers of neurons and have limited performance in complex tasks compared to state-of-the-art electronic artificial neural networks (ANNs). Diffractive photonic computing \cite{lin2018all, zhou2021large, kulce2021all} supports millions of neurons and arbitrary linear transformation, which can be utilized to construct more advanced architecture to perform complex AI tasks. We demonstrate the use of diffractive photonic computing to construct diffractive PNN architectures and perform the EEG analysis for epileptic seizure detection.

Epilepsy is a serious chronic neurological disorder that is repetitive and sudden. It has been proved that approximately one-third of the patients are intractable with medications \cite{litt2002prediction}. They are at high risk of serious physical injury or even death if suffering accidental seizures. Therefore, performing the EEG analysis to achieve automated and portable seizure detection devices are critical to alert caregivers and protect the lives of patients. Many seizure detection methods based on statistical features and machine learning classifiers have been proposed \cite{shoeb2010application, siddiqui2020review}, where the ANN-based methods \cite{zhou2018epileptic, daoud2019efficient} have achieved state-of-the-art detection performance. Instead of executing ANNs energy intensively on traditional electronic edge devices with limited computing power, PNNs can overcome the limitations of electronics \cite{zhang2019artificial} and provide promising opportunities to design portable and power-saving seizure detection devices, which has not been demonstrated.

Here, we propose the EEG opto-processor based on diffractive photonic computing units (DPUs) and construct the PNN architecture to perform the epileptic seizure detection with EEG and iEEG measurements (Fig. \ref{Fig1}). We develop the 3D free-space DPU with a highly compact optical system, where a high data-throughput spatial light modulator (SLM) is used to implement the input data encoding and the diffractive modulation simultaneously. The diffractive deep neural networks ($\rm D^2NNs$) are constructed with DPU by implementing the computational operations of diffractive weighted interconnections optically, using photoelectric conversion for nonlinearity activation, and controlling the dataflow electronically. Furthermore, we design a 2D integrated DPU with a metaline-based multi-layer structure \cite{doi:10.1126/sciadv.abn7630, wang2022integrated, wang2019chip} on a silicon photonics platform to achieve higher integration density and lower power consumption towards wearable devices. In this design, we further introduce an optical bias block and validate its effectiveness in unbalanced classification tasks in this work. We demonstrate that the DPU-based $\rm D^2NNs$ successfully processes the multi-channel EEG and iEEG signals and detects the epileptic seizure with high performance. Besides, channel selection using random forest is developed to select the most important channels. The results show that the selected single-channel signal can be used for the detection with sufficient detection accuracy. We conduct the evaluations theoretically and experimentally, providing the possibility to collect EEG and iEEG signals with only one electrode for each patient and making the system more economical, efficient, and comfortable for personalized medicine.

\section{Results}\label{sec2}

\subsection{Epileptic seizure detection with the free-space DPU}\label{subsec2}

\begin{figure}[h]
\centering
\includegraphics[width=1\textwidth]{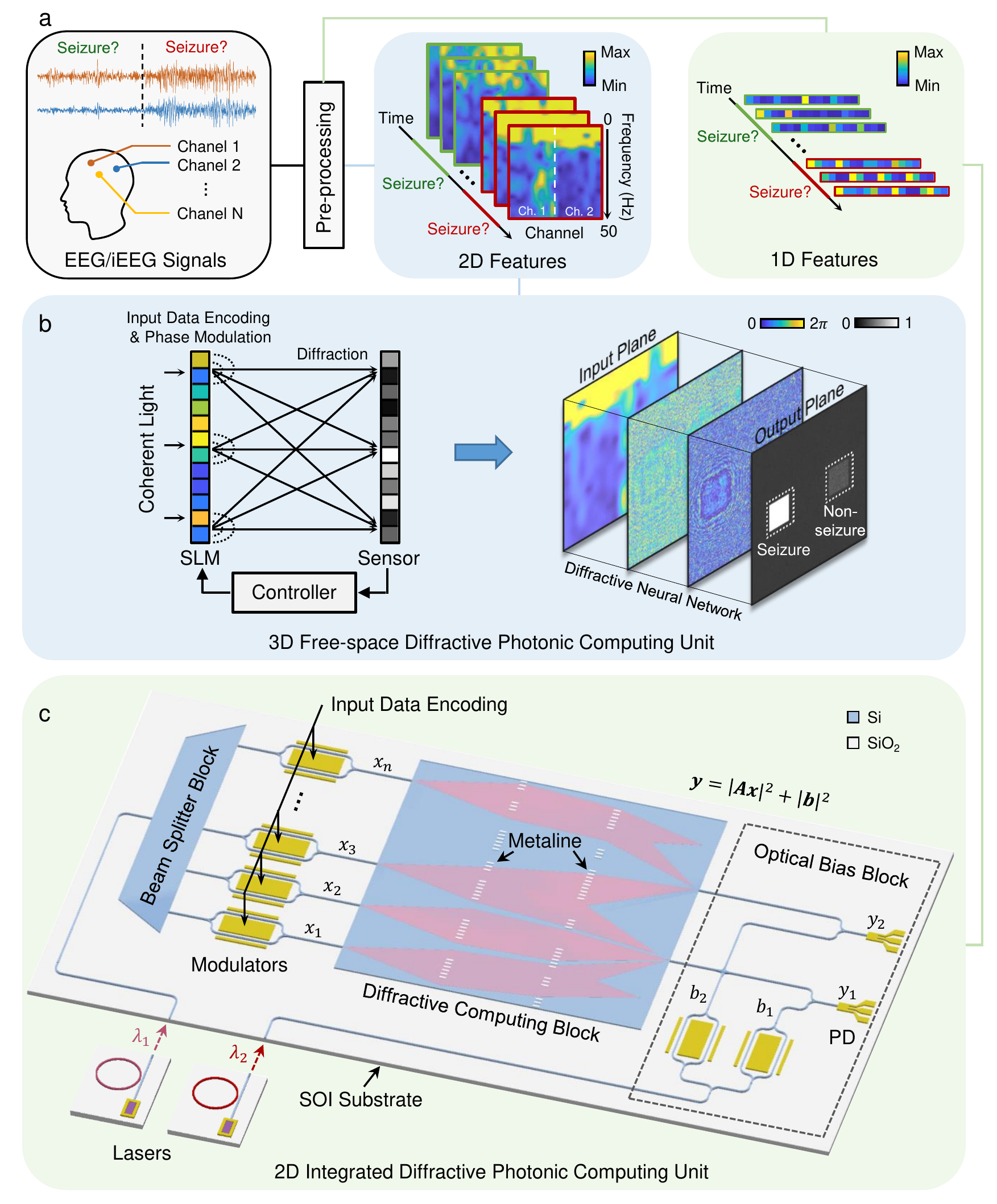}
\caption{The architecture of EEG opto-processor. \textbf{a}, The EEG and iEEG channel signals are pre-processed to extract the 2D or 1D statistical features through short-time Fourier transform (STFT) for epileptic seizure detection. \textbf{b}, The 3D free-space DPU comprises an SLM for input data encoding and phase modulation, an optoelectronic detector for nonlinear activation and capturing diffractive computation results, and an electronic controller to configure the dataflow. With the constructed $\rm D^2NNs$ \cite{lin2018all} using DPU, the seizure detection result is determined by the light intensity distribution of two target regions on the output plane. \textbf{c}, A schematic of the integrated DPU on an SOI platform. Input data are encoded in the amplitude of the light in optical waveguides with modulators, weighted interconnected through the diffractive modulation of metalines, and biased through incoherent energy coupling and photoelectric conversion.}
\label{Fig1}
\end{figure}

We utilize photonic computing to analyze and detect the epileptic seizures from the recorded electroencephalogram (EEG) signals (see Fig. \ref{Fig1}a). EEG is one of the most used methods for monitoring brain activity and is considered the best indicator for epilepsy diagnosis and analysis \cite{french2008initial}. The extracranial EEG measures the voltage difference between scalp-mounted electrodes caused by the ionic flowing in the brain neurons, while the intracranial electroencephalogram (iEEG) is the neuroelectrophysiologic signal obtained from implanted electrodes. Different channels from the electrodes at the corresponding positions reflect the spatial and temporal information of brain activity. The epileptiform EEG or iEEG patterns, such as spikes and sharp waves, contribute to the seizure diagnosis. The effectiveness of the proposed EEG opto-processor is validated on both EEG and iEEG modalities.

We develop a 3D free-space DPU to execute the task by pre-processing the EEG signals into 2D images to be analyzed with the constructed large-scale $\rm D^2NNs$ architecture \cite{zhou2021large}, as shown in Fig. \ref{Fig1}b. The main components are the large-scale reconfigurable SLM and the optoelectronic detector that can be programmed to support millions of diffractive neurons. The input data are encoded on the phase of the optical field to the input nodes and modulated by the diffractive layer parameters that can be trained according to specific AI tasks. The input nodes are connected to output neurons via optical diffraction, and the trainable synaptic weights are generated by the diffractive modulation. The nonlinear activation function occurs in the photoelectronic conversion of the diffractive computation results by measuring the intensity of complex fields and intensity-to-phase conversion at the next input layer. The dataflow is controlled electronically for building the multi-layer PNN architecture that performs video-rate epileptic seizure detection. Different from \cite{zhou2021large}, both the input data encoding and phase modulation are implemented with SLM, which eliminates the additional input data encoding module and the related relay optics for significantly reducing the system complexity. The experimental setup of the free-space DPU is depicted in Supplementary Fig. 1. We construct a 2-layer $\rm D^2NN$ with the modeling and training process detailed in \textit{Materials and Methods}. For each input signal sequence, two target regions, representing the seizure and non-seizure states, are configured at the network output plane, where the target region with higher optical intensity indicates the epileptic seizure detection result. After the training to learn the diffractive modulation coefficients, the $\rm D^2NN$ performs light-speed inference, offering video-rate and energy-efficient epileptic seizure detection for patients.

We first validate the effectiveness of the proposed optical epileptic seizure detection approach using EEG signals of the CHB-MIT dataset \cite{shoeb2009application, goldberger2000physiobank} from Children Hospital Boston, Massachusetts Institute of Technology. It contains hours of seizure and non-seizure EEG recordings of 23 patients obtained through the 10-20 International system of EEG electrode placement. The CHB-MIT dataset is widely used as a benchmark for epileptic seizure detection \cite{birjandtalab2017automated, zhou2018epileptic, daoud2019efficient, siddiqui2020review}. Similar to \cite{daoud2019efficient}, we selected the EEG recordings of 8 patients with sufficient seizure time (more than 400 seconds) to generate the training and testing dataset for the $\rm D^2NN$ (see Methods, and Supplementary Table 1). The EEG signals were pre-processed and transformed into images to match the input format of the $\rm D^2NN$. Besides, since the epileptic EEG signals are nonstationary and multi-component, extracting meaningful statistical features from raw 1D timing signals is critical for the detection. We adopted the short-time Fourier transform (STFT) to pre-process the raw signals and extract effective 2D features containing both time and frequency characteristics \cite{tzallas2009epileptic, boonyakitanont2020review}. STFT can be easily implemented in wearable devices based on digital electronics, and optical solutions have also been demonstrated to achieve higher speed and bandwidths \cite{li2011all, xie2020stft}. Epileptic seizures will cause changes in certain frequency bands, such as $\delta$ (0.4–4 Hz), $\theta$ (4–8 Hz), $\alpha$ (8–12 Hz), $\beta$ (12–30 Hz), and $\gamma$ (30–70 Hz) bands \cite{tzallas2009epileptic, tatum2014ellen}. And the energy of the EEG signals is mainly concentrated in low-frequency bands, so we set the frequency range of STFT to 0-50 Hz. For each time window of EEG signals, the signals of different channels were transformed into STFT spectra, stitched together into one image, and fed to a $\rm D^2NN$ to detect epilepsy seizures.

Selecting informative EEG channels is critical for building wearable healthcare devices and achieving personalized medicine in epileptic seizure detection. Generalized epileptic seizures involving the whole brain can be seen in every channel of the EEG recordings, while partial seizures can be seen only in a few channels whose corresponding electrodes are near the lesion. Full-channel EEG monitoring confronts great challenges in wearable devices because it is expensive, time-consuming, computationally intensive, and uncomfortable. Moreover, invalid channels add noise to the signals and increase detection difficulties. Channel selection can be achieved by medically diagnosing the location of the lesion or with statistical methods \cite{birjandtalab2017automated, alotaiby2015review}. This process only needs to be completed in the preliminary analysis but not repeated in the subsequent epilepsy monitoring.

\begin{figure}[h]
\centering
\includegraphics[width=1\textwidth]{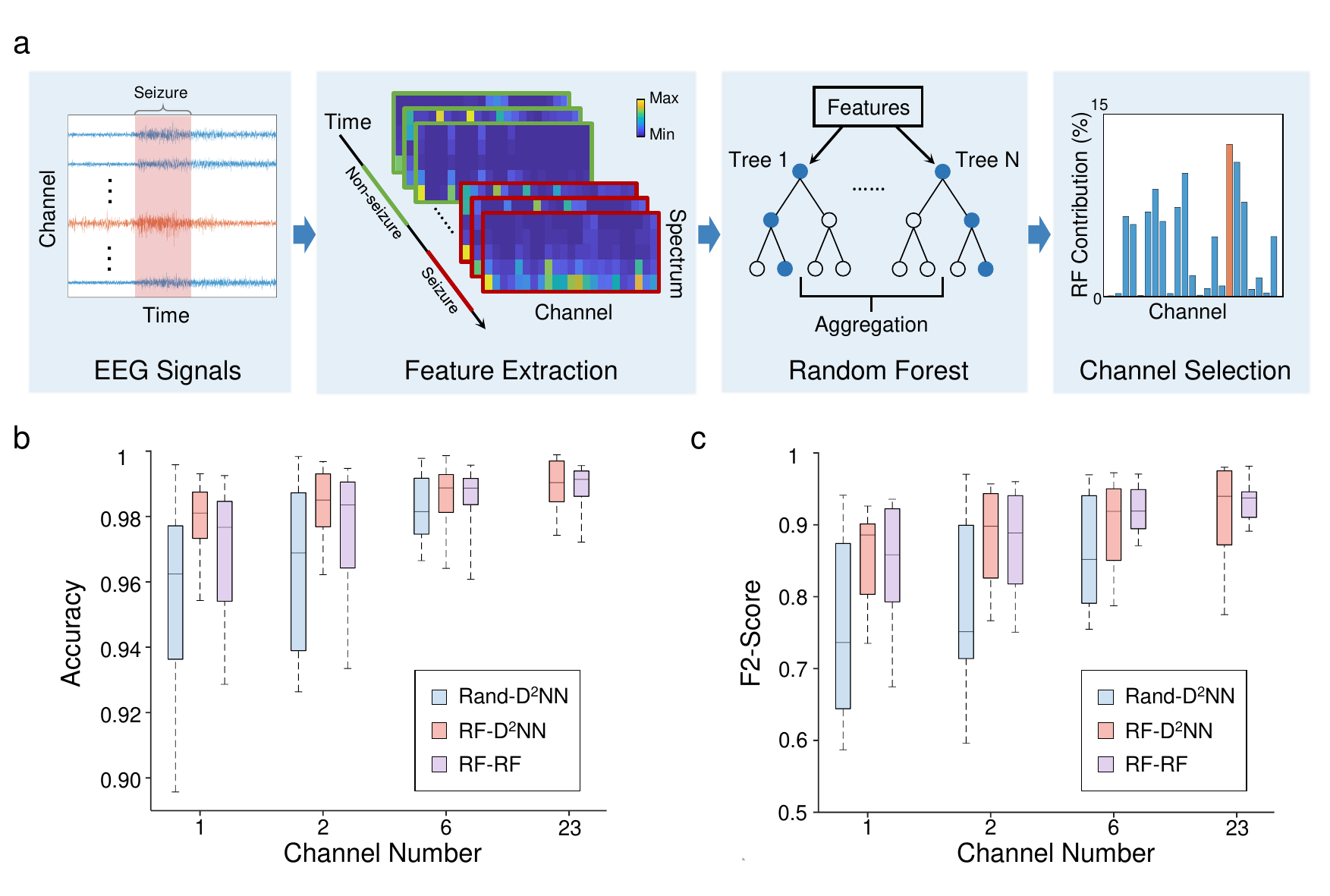}
\caption{Numerical evaluations of free-space DPU on the CHB-MIT dataset under different EEG channel numbers. \textbf{a}, The pipeline of channel selection using random forest (RF). The EEG signals are segmented into sub-sequences given the time intervals, and the power spectral density of each channel is calculated as five features, representing $\delta$, $\theta$, $\alpha$, $\beta$, and $\gamma$ bands, for learning the random forest model. The feature contribution percentage is the normalized total reduction of the impurities in the random forest brought by that feature, i.e., the Gini importance, and the channel contribution percentage is the sum of five feature contribution percentages. The channels with the highest contribution percentage are selected for limited-channel seizure detection. \textbf{b} and \textbf{c}, The box diagrams of the classification accuracies and F2 scores obtained by blindly testing on the CHB-MIT dataset under different channel numbers. Rand-$\rm D^2NN$, RF-$\rm D^2NN$, and RF-RF denote selecting channels randomly and classifying with the $\rm D^2NN$, selecting channels with a random forest and classifying with $\rm D^2NN$, and selecting channels and classifying simultaneously with a random forest, respectively.}
\label{Fig2}
\end{figure}

We apply a random forest-based channel selection method \cite{birjandtalab2017automated} to analyze the performance of the $\rm D^2NN$ for EEG analysis on the CHB-MIT dataset under different channel numbers (see Fig. \ref{Fig2}). The pipeline of channel selection is shown in Fig. \ref{Fig2}a, where the EEG signals are first divided into sub-sequences with an equivalent time interval of 1-second. Next, the power spectral density of each EEG channel at each window is calculated, representing different bands, i.e., $\delta$, $\theta$, $\alpha$, $\beta$, and $\gamma$. The signals of each time window are utilized as a sample with 115 attributes (23 channels and 5 bands). A random forest \cite{breiman2001random} containing 1000 decision trees is used to learn the features of each sample and detect the epileptic seizures, which has been proven to be robust in learning irrelevant features \cite{kursa2014robustness}. During the process of updating the random forest, i.e., minimizing the sum of the impurities, redundant features will not be selected for node splitting, and informative features will appear in those trees more frequently. Therefore, the feature contribution percentage can be obtained by investigating the normalized total reduction of the impurities in the random forest brought by that feature, which is also known as the Gini importance. And the channel contribution percentage is the sum of five feature contribution percentages. For example, the rightmost subgraph of Fig. \ref{Fig2}a illustrates the contribution percentage of 23 channels for the patient chb01 (see Supplementary Table 1), and the most important channel is 17. The channels with the highest contribution percentage are selected for seizure detection with $\rm D^2NN$. In addition, the random forest is also an effective machine learning classifier for epileptic seizure detection \cite{siddiqui2020review} and serves as a performance reference for the $\rm D^2NN$-based approach.


With the channel selection, we evaluate the performance of $\rm D^2NN$ on the CHB-MIT dataset. The statistical evaluations, including accuracy, sensitivity, specificity, and $F_{\beta}$ score, are used for evaluating the binary classification performance. In the medical domain, the $F_{\beta}$ score, with typically $\beta=0.5, 1, 2$, is a weighted harmonic mean of the precision and recall that is more useful than accuracy, especially when the dataset is imbalanced \cite{powers2020evaluation}. We adopt the F2 score, i.e., $\beta=2$, to allow the recall to have a larger weight, considering the identification of disease episodes with only a minority of time \cite{devarriya2020unbalanced}. We explore the performance of $\rm D^2NN$ under different layer numbers with the 1-channel signals selected by random forest (see Supplementary Fig. 2). With 0.16 million diffractive neurons in each layer, the average accuracy, sensitivity, specificity, and F2 score are 97.89\%, 91.92\%, 98.13\%, and 0.8553, respectively, for the 2-layer $\rm D^2NN$, which are comparable to the 3-layer model. Further increasing the layer number has a little performance improvement. Besides, the 1-layer $\rm D^2NN$ does not converge during training and is not competent for this task. The same conclusion was obtained for larger channel numbers. Therefore, the 2-layer $\rm D^2NN$ model is adopted in the numerical evaluations and experiments.

The accuracy and F2 score of the $\rm D^2NN$ numerical model under different channel numbers are visualized with the box diagram in Fig. \ref{Fig2}b and Fig. \ref{Fig2}c, respectively. Rand-$\rm D^2NN$, RF-$\rm D^2NN$, and RF-RF denote selecting channels randomly and classifying with a $\rm D^2NN$, selecting channels with a random forest and classifying with a $\rm D^2NN$, and selecting channels and classifying simultaneously with a random forest, respectively. Both the accuracy and F2 score of RF-$\rm D^2NN$ outperform Rand-$\rm D^2NN$ and RF-RF obviously when using a few channels, demonstrating the effectiveness of the channel selection method using random forest. The performances of all models improve with the increase of the channel number, especially for the Rand-$\rm D^2NN$ and RF-RF models. The RF-$\rm D^2NN$ model achieves sufficient high performance using 1-channel EEG signals with an average accuracy of 97.89\% and an F2 score of 0.8553, which are close to the full-channel RF-$\rm D^2NN$ model with an average accuracy of 98.96\% and an F2 score of 0.9161. Notice that the full-channel RF-RF achieves an average accuracy of 98.88\% and an F2 score of 0.9335. Besides, the 1-channel RF-$\rm D^2NN$ performs relatively stable on different patients, i.e., the accuracy varies from 95.43\% to 99.31\%, and the F2 score varies from 0.7351 to 0.9261. The same conclusion was obtained from the sensitivity and specificity metrics (see Supplementary Fig. 3). The simulation results verify that the $\rm D^2NN$ with random forest for channel selection has the sufficient ability to make full use of the information of 1-channel signals to perform high-quality epileptic seizure detection.

\begin{figure}[h]
\centering
\includegraphics[width=1\textwidth]{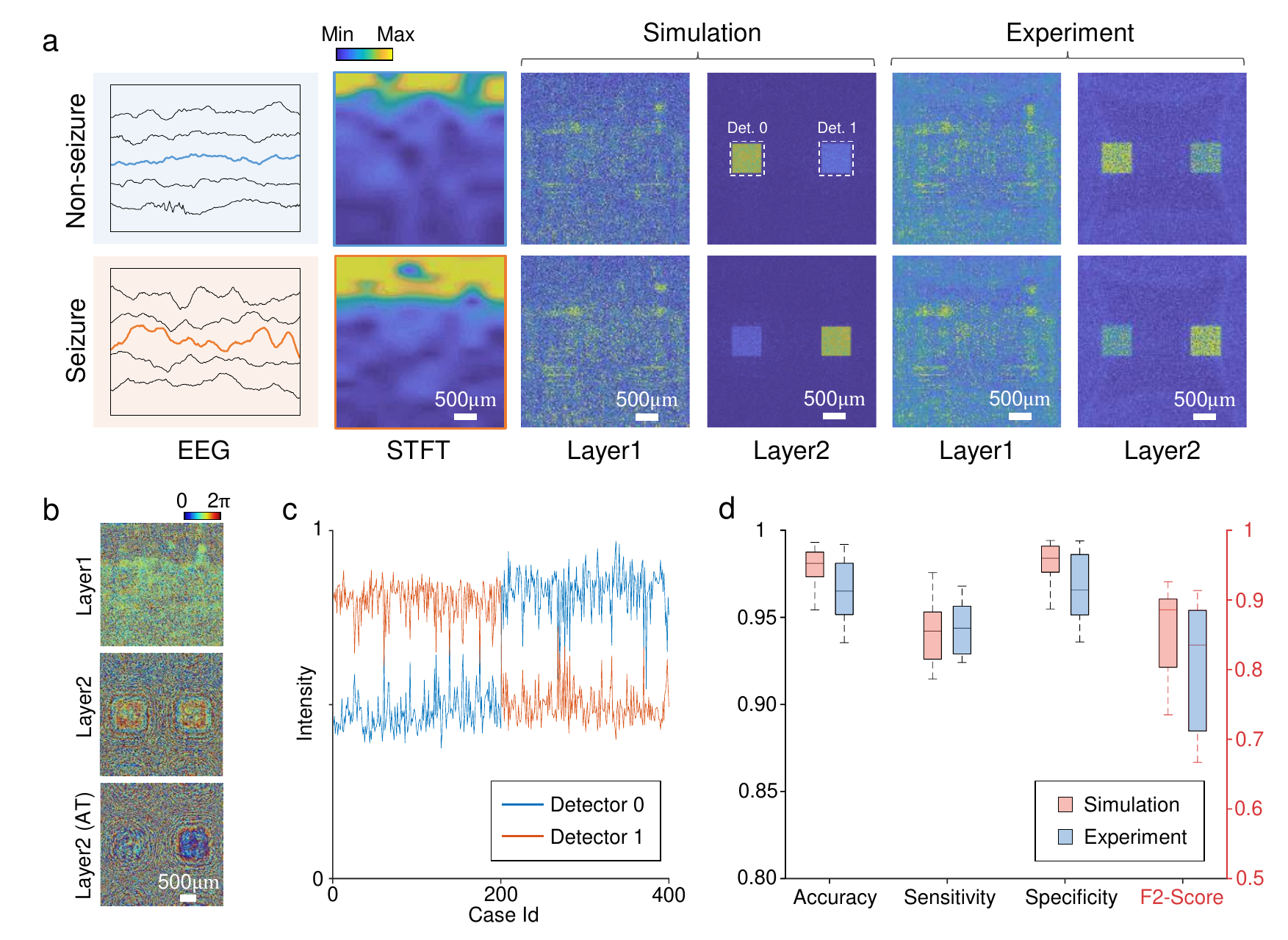}
\caption{Experimental results of the free-space DPU on the CHB-MIT dataset. \textbf{a}, The simulation and experiment outputs of 2-layer $\rm D^2NNs$ on the patient chb01 (see Supplementary Table 1) using 1-channel EEG signals with the top channel contribution percentage. \textbf{b}, The phase modulation coefficients of the pre-trained $\rm D^2NNs$ model and the fine-tuned second-layer phase modulation coefficients after adaptive training (AT). \textbf{c}, The experimental measured the average intensity of two detector regions with instances of 200 seizures and 200 non-seizures. \textbf{d}, The accuracy, sensitivity, specificity, and F2 score of simulations and experiments on all the patients.}
\label{Fig3}
\end{figure}

We conduct experiments using the 2-layer model and single-channel EEG signals to facilitate more efficient computing and convenient signal acquisition. The results are presented in Fig. \ref{Fig3}. In order to eliminate the systematic errors, the modulation coefficients of the second layer are fine-tuned via adaptive training with the experimental results of the first layer (see Methods). Fig. \ref{Fig3}b illustrates the pre-trained parameters of the 2-layer $\rm D^2NN$ and the fine-tuned parameters of the second layer. Additionally, the intensities of the two target detection regions are multiplied by two factors, one for each region, to reduce the impact of uneven illumination and improve detection accuracy (see Methods). Fig. \ref{Fig3}a illustrates the experimental results of the $\rm D^2NNs$ on the patient chb01 using 1-channel EEG signals. As expected, the $\rm D^2NNs$ successfully distinguish the seizure and non-seizure samples and concentrate the light into predefined regions. And the image characteristics of the experimental results are very close to the simulation results. Fig. \ref{Fig3}c demonstrates the normalized intensity of 2 detector regions for 200 seizures and 200 non-seizures. The distinct difference makes the system robust for small noise disturbances. Furthermore, we evaluate the numerical and experimental performances of the 1-channel model on all the patients. As shown in Fig. \ref{Fig3}d, the average accuracy, sensitivity, specificity, and F2 score are 96.84\%, 92.81\%, 96.98\%, and 0.8174, respectively, in the experiments, which are comparable with the simulation. Detailed results for each patient are shown in Supplementary Table 2, verifying the correctness and effectiveness of the 1-channel approach and the experimental setup.

\begin{figure}[h]
\centering
\includegraphics[width=1\textwidth]{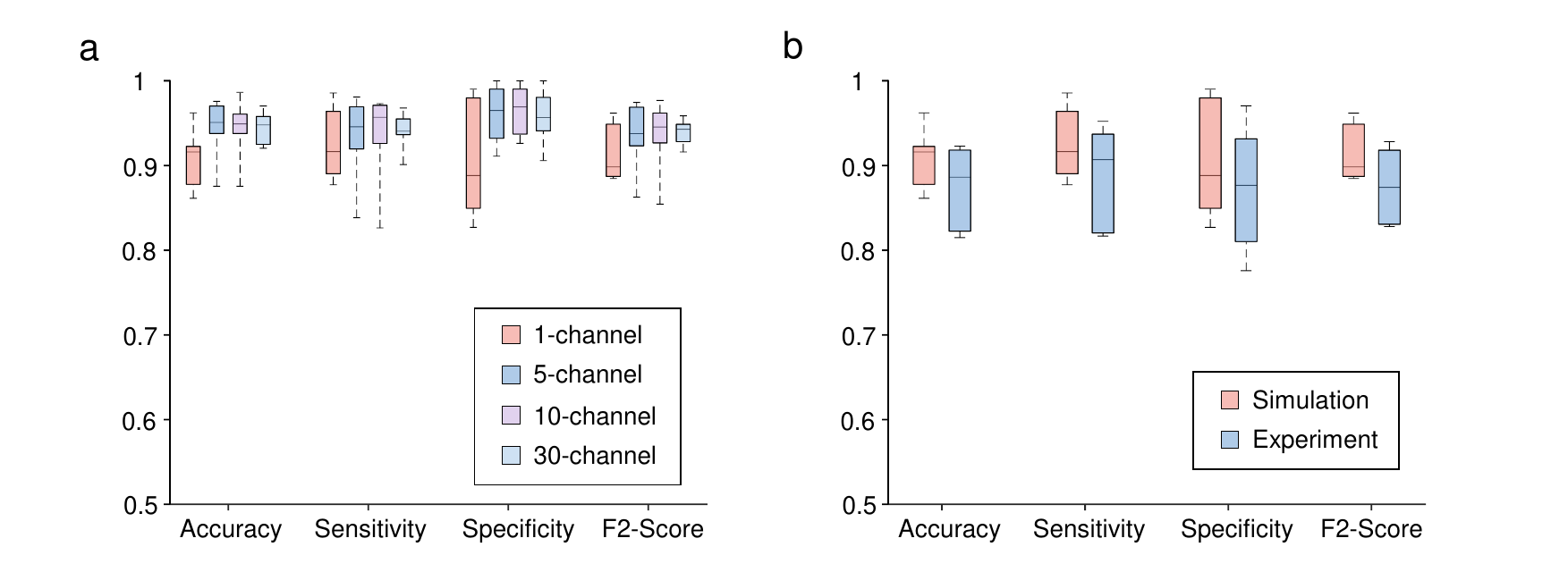}
\caption{Performance of epileptic seizure detection with iEEG signals on the Epilepsy-iEEG-Multicenter-Dataset dataset. \textbf{a}, The numerical evaluation results under different channel numbers. \textbf{b}, The performance comparison between numerical evaluations and experiment results using the single-channel iEEG signals.}
\label{Fig4}
\end{figure}

In addition to the EEG modality, we further utilize the DPU-constructed $\rm D^2NN$ to detect epileptic seizures with iEEG signals. The performance is evaluated on electrocorticography (ECoG) recordings from 6 patients selected from the Epilepsy-iEEG-Multicenter-Dataset \cite{ds003029:1.0.3} (see Methods, Supplementary Table 1 and Table 2). Each iEEG signal sequence of these patients has a time duration of a few minutes near an epileptic seizure with large numbers of seizure signals, resulting in more difficulties for the detection compared with the CHB-MIT dataset that contains a long duration of non-seizure periods. We adopt the same feature exaction, channel selection method, and $\rm D^2NN$ model as in EEG signals, and the performance under different channel numbers is shown in Fig. \ref{Fig4}a. With the channel number settings of 1, 5, 10, and 30, the average accuracy is 90.93\%, 94.35\%, 94.31\%, and 94.49\%, respectively, and the average sensitivity is 92.50\%, 93.32\%, 93.51\%, and 94.03\%, respectively; the average specificity is 90.38\%, 96.06\%, 96.53\%, and 95.68\%, respectively, and the average F2 score is 0.9132, 0.9341, 0.9351, and 0.9395, respectively. The model's generalization ability on different patients is further demonstrated in Supplementary Fig. 4. Similar to detecting epileptic seizure with EEG signals, we adopt the single-channel iEEG signals after channel selection for the experiments, as shown in Fig. \ref{Fig4}b. The average accuracy, sensitivity, specificity, and F2 score are 87.51\%, 89.00\%, 87.35\%, and 0.8755, respectively, in the experiments.

\subsection{Epileptic seizure detection with the integrated DPU}\label{subsubsec2}

We propose to design a 2D integrated DPU based on metamaterials \cite{doi:10.1126/sciadv.abn7630, wang2022integrated} and on-chip optical devices for further reducing the system size and improving the computational efficiency towards the wearable EEG analysis, as shown in Fig. \ref{Fig1}c. Specifically, the coherent light from an on-chip laser \cite{zhou2015chip} is split into different single-mode waveguides with an array of multi-mode interferometers (MMIs) or Y-couplers. The 1D features of EEG signals are encoded on the light amplitude in the waveguides with on-chip modulators, such as Mach-Zehnder interferometers (MZIs). The weighted interconnections between the complex optical field of input and output waveguides are achieved with a diffractive computing block comprising the multi-layer metalines, which are passive optical structures for high energy efficiency computations. Each metaline is a 1D etched rectangle silica slot (meta-atom) array in the silicon membrane of the silicon-on-insulator (SOI) substrate. Both amplitude and phase modulation coefficients of a meta-atom are learnable and can be programmed by adjusting the width and length of the slot, where we train only the width in this work (see Supplementary Fig. 5). The diffractive computing block can be expanded vertically to receive higher dimensional input features and horizontally to increase diffractive parameters and learning capacities. In addition to the weighted interconnection block, we introduce an optical bias block to enhance the model capability. The bias coefficients are modulated at a different wavelength and coupled into two output waveguides of the diffractive computing block. The incoherent summation of optical energy is collected with photodetectors (PD) to indicate the epileptic seizure detection results. The bias modules enable the adjustment of output thresholds, which is especially effective in imbalanced data, such as epileptic seizure signals used in this work \cite{zhou2005training}.

\begin{figure}[h]
\centering
\includegraphics[width=1\textwidth]{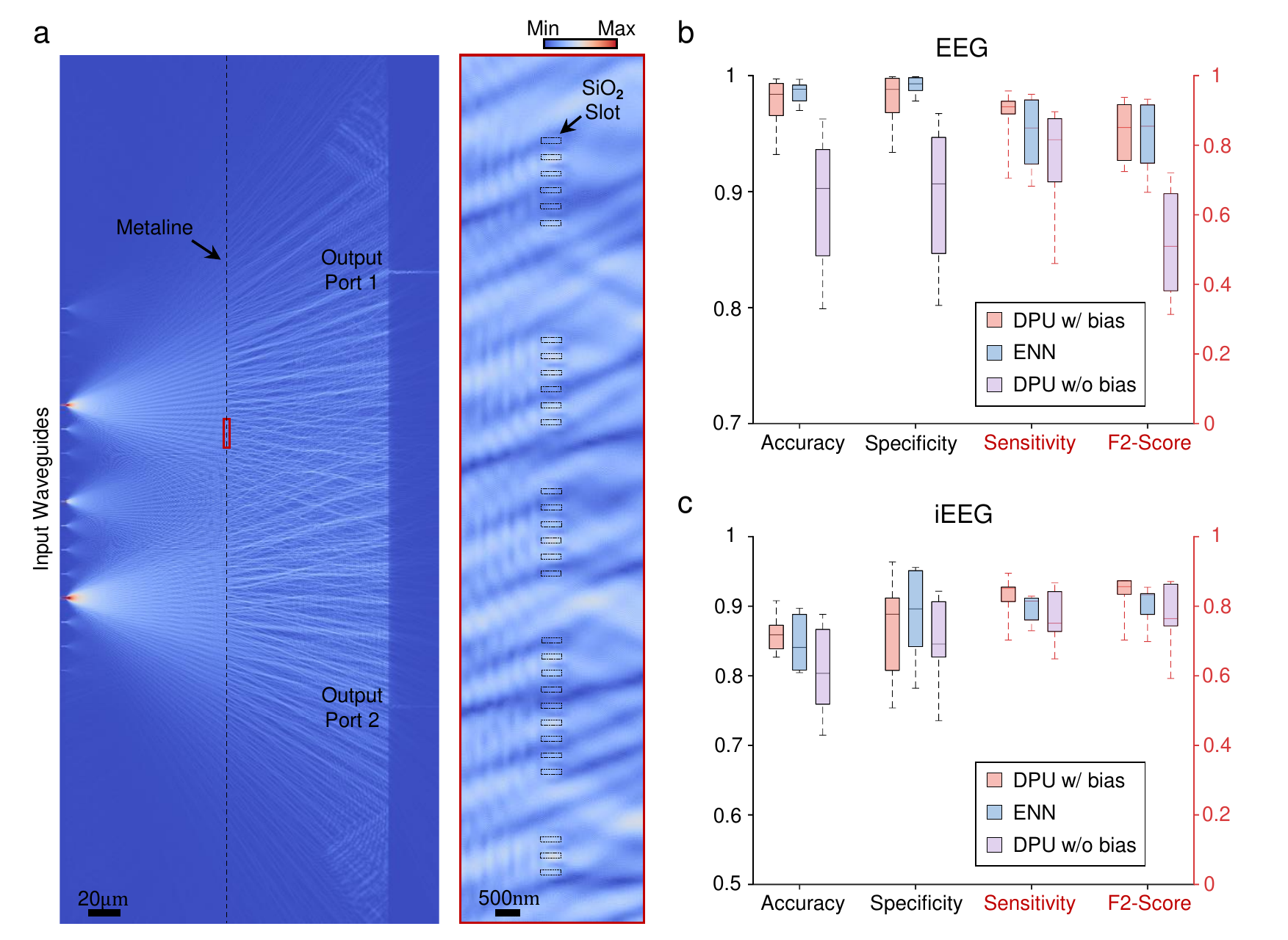}
\caption{Epileptic seizure detection with the integrated DPU. \textbf{a}, The diffractive optical field propagation of the integrated DPU simulated with FDTD. The right subgraph shows the enlarged view of the red box at the position of metaline in the left subgraph. \textbf{b} and \textbf{c}, Results on the EEG and iEEG signals from the CHB-MIT dataset and the Epilepsy-iEEG-Multicenter-Dataset, respectively. The performance of the integrated DPU with the optical bias block is comparable with the electronic neural network (ENN) but reduced obviously without the optical bias block.}
\label{Fig5}
\end{figure}

We apply the integrated DPU for epileptic seizure detection on EEG signals from the CHB-MIT dataset. To reduce the input port numbers of DPU, the time and frequency resolution of STFT is reduced during the pre-processing and only one channel selected by random forest is utilized to generate a 16-dimensional feature for each time window (see Methods). Since the modulation coefficients of each meta-atom are calculated with the periodic boundary condition by finite-difference time-domain (FDTD) simulations, we use binary modulation and combine three identical meta-atoms as one diffractive modulation neuron to facilitate the fabrication and improve the model accuracy. We trained the integrated DPU with 16 input waveguides and a metaline with 600 diffractive neurons considering the module size. Fig. \ref{Fig5}a illustrates the optical field propagation details simulated with FDTD. The sufficient matched modulation profiles between the analytical model and FDTD are demonstrated in Supplementary Fig. 6. We numerically evaluate the integrated DPU with an optical bias block in Fig. \ref{Fig5}b and Supplementary Table 3. The average accuracy, sensitivity, specificity, and F2 score are 97.68\%, 88.92\%, 98.01\%, and 0.8385, respectively, but reduced to 89.11\%, 76.61\%, 89.61\%, and 0.5173, respectively, without the optical bias block. Besides, the performance fluctuates greatly in different patients without the optical bias block. Notice that the average accuracy, sensitivity, specificity, and F2 score of an electronic fully connected neural network (ENN) with the same number of parameters are 98.54\%, 83.49\%, 99.17\%, and 0.8292, respectively. The results demonstrate the effectiveness of the optical bias block for integrated DPU and the success of its application for epileptic seizure detection.

Similarly, the results on iEEG signals from the Epilepsy-iEEG-Multicenter-Dataset are shown in Fig. \ref{Fig5}c and Supplementary Table 3. The average accuracy, sensitivity, specificity, and F2 score are 86.07\%, 82.82\%, 86.90\%, and 0.8328, respectively, but reduced to 80.60\%, 76.45\%, 84.70\%, and 0.7665, respectively, without the optical bias block. Compared to the CHB-MIT dataset, there is less performance degradation because the duration of seizure and non-seizure periods are similar in this dataset. The ENN with the same number of parameters achieves an average accuracy, sensitivity, specificity, and F2 score of 84.66\%, 79.54\%, 88.72\%, and 0.8051, respectively, also close to the integrated DPU with an optical bias block.

\section{Discussion}
\subsection*{Computing speed and energy efficiency}
We evaluate the computing performance of the free-space DPU with the constructed 2-layer $\rm D^2NN$. Each layer of $\rm D^2NN$ generates a feature map with the element numbers of $400\times400$ from an input image with the same size, which performs $5.12\times10^{10}$ operations in the optical domain. Since the diffractive optical computations are operated at the speed of light, the computing speed is determined by the refresh rate of the SLM (30 Hz) and the acquisition frame rate of the CMOS optoelectronic detector (60 Hz). Considering a 30 Hz system frame rate, the computing speed of free-space DPU is $1.536\times10^{12}$ operations per second (1.536 TOPS). Furthermore, the free-space DPU system supports $1920\times1152$ diffractive neurons determined by the SLM. As a result, the maximum diffractive computation speed is 293.53 TOPS, which is better than the state-of-the-art edge-computing device of Nvidia Jetson AGX Xavier with a computing speed of 32 TOPS.

The integrated DPU in Fig. \ref{Fig5}a executes a $16\times2$ weight matrix with passive metalines in a region of $540~\rm\mu m \times200~\rm\mu m$. Considering the typical light source power of 10 mW with a modulation and photodetection rate of 30 GHz based on the existing silicon photonic foundry, the computing speed, computing density, and energy efficiency are 1.92 TOPS, 17.778 TOPS per square millimeter and 192 TOPS per watt, respectively. Notice that the computing density and energy efficiency of the state-of-the-art GPU Tesla V100 are 0.037 TOPS per square millimeter and 0.1 TOPS per watt, respectively \cite{yao2020fully}. The integrated DPU achieves more than two orders of magnitude improvements in computing density and more than three orders of magnitude improvements in energy efficiency.

\subsection*{Limitations and future works}
We have demonstrated the DPU-based opto-processor for effectively processing large-scale EEG signals, which can facilitate its applications in different areas. For example, in the brain-computer interface \cite{even2020power}, the iEEG signal channel numbers are increasing from thousands to hundreds of thousand, placing high demand for high-performance computing processors. Furthermore, we have validated that the 1-channel EEG signals with the channel selection method have sufficiently high performance for epileptic seizure detection, i.e., only one electrode attached to the scalp is required for the acquisition system, facilitating its applications in wearable healthcare and personalized medicine. The volume of the free-space DPU can be smaller by designing the application-specific integrated circuits for EEG signal acquisition and dataflow control.

The reconfigurable free-space DPU includes millions of diffractive neurons, which can construct more complex neural networks architectures, such as the diffractive recurrent neural network \cite{zhou2021large} for better performance in processing time series signals and different health monitoring tasks. The integrated implementation can be fabricated with deep ultraviolet lithography, which is highly compatible with the electronic integrated circuits. Reconfigurable non-volatile materials, such as phase change materials (PCM) \cite{feldmann2021parallel, wu2021programmable}, can be used to design the metalines with programmable network parameters and increase the flexibility. Although the 1-channel model is mainly considered in this article, more channels may be necessary for more complex tasks. The data throughput of the proposed EEG opto-processor can be further expanded based on wavelength division multiplexing (WDM) \cite{feldmann2021parallel, xu202111}, and the optical processing of different channels can be accomplished at individual wavelengths.

In summary, we have demonstrated the successful use of DPU to construct photonic neural networks for EEG analysis and apply it for the epileptic seizure detection of EEG and iEEG signals with high performance. This work will inspire promising advances in using photonic computing for developing various healthcare devices.

\section{Methods}\label{sec11}

\subsection*{Data preprocessing}
The EEG recordings of 8 patients, each with sufficient length of seizure time duration (more than 400 seconds) and uniformly recording of 23 channels, were selected from the CHB-MIT dataset. The signals of seizures and randomly selected 2-hour non-seizures were used for evaluations. All the EEG signals were segmented into 1-second windows and randomly arranged. Half of the seizures and the non-seizures with the same duration were used for training, while the remaining data were used for testing. The channels of EEG signals were selected according to the random forest-based method, and the STFT was adopted to extract time-frequency-domain features for each time window. For the evaluation of free-space DPU, the frequency range of the STFT was set to 0-50 Hz, and the sliding window width was set to 51 samples. The spectral energy was normalized, and the 2D features were resized to $400\times400$ pixels as the input of $\rm D^2NNs$. For the evaluation of integrated DPU, each time window of 1-second signals was divided into four parts, and the energy of four spectral bands (0-6 Hz, 6-14 Hz, 14-22 Hz, 22-30 Hz) in each part were calculated to form a 1D feature with 16 attributes.

Similarly, the iEEG recordings of 6 patients with sufficient length of seizure time duration and clear onset and offset labels were selected from the Epilepsy-iEEG-Multicenter-Dataset. They were segmented using 5-second sliding windows with an overlap of 4 seconds and were further randomly arranged. The training and testing sets were divided in the same way as the EEG recordings. For the evaluation of free-space DPU, the frequency range of the STFT was set to 0-100 Hz due to the higher sampling frequency of the iEEG signals, and the sliding window width was set to half of the sampling frequency. For the evaluation of integrated DPU, each time window of 5-second signals was divided into five parts and four spectral bands, and a 1D feature with 20 attributes was generated.

\subsection*{Experimental system of the free-space DPU}
The experimental system is shown in Fig. \ref{Fig1}b and Supplementary Fig. 1. A compact green LED (CPS532, Thorlabs) coherent light source at the wavelength of 532 nm was used to generate the input optical field. The light beam is collimated and expanded by two lenses (AC050-008 and AC254-100, Thorlabs), polarized with a polarizer (LPVISA050, Thorlabs), and split with a beam splitter (CCM1-BS013, Thorlabs). An SLM (P1920-400-800, Meadowlark) modulates the phase of the wavefront according to the input information and the trained parameters of the $\rm D^2NN$. The optical field is captured and photoelectrically converted with a complementary metal–oxide–semiconductor (CMOS) sensor (GS3-U3-41C6M-C, Point Grey) after diffraction propagation of 10 cm. This process is multiplexed and programmed to construct an $\textsl{N}$-layer $\rm D^2NN$.

\subsection*{Modeling of the free-space DPU}
The diffractive computing process can be represented as $\textbf{y}_i={\mid F\cdot {\rm exp}(j(\textbf{x}_i+\textbf{H}_i))\mid}^2$, where $\textbf{x}_i$, $\textbf{H}_i$, $\textbf{F}$, and $\textbf{y}_i$ denote the input, the trainable modulation coefficients, the transmission matrix of the free-space diffractive propagation, and the output of the $i$-th layer of the $\rm D^2NN$, respectively. $\textbf{x}_i+\textbf{H}_i$ is fed to the SLM to modulate the phase of the light wavefront and $\textbf{y}_i$ is the light intensity acquired by the CMOS sensor. $\textbf{x}_1$ is the 2D feature map scaled to [0, 2$\pi$] of the EEG or iEEG signals and $\textbf{x}_i$ is obtained from $\textbf{y}_{i-1}$ for $i>1$ via a nonlinear activation function, given as $\textbf{x}_i=2\pi \cdot{\rm Sigmoid}(a_i\cdot \textbf{y}_{i-1}+b_i)$, where $a_i$ and $b_i$ are two trainable parameters. Finally, $\textbf{y}_N$ represents the output plane of the $\rm D^2NN$, and the target region with higher optical intensity indicates the epileptic seizure detection result. Supplementary Fig. 1 demonstrates the diffractive computing process of a 2-layer $\rm D^2NN$. The network diffractive modulation coefficients are trained with the pre-collected and labeled data. The mean squared error (MSE) between $\textbf{y}_N$ and the ground truth, i.e., 1 for the target detection region and 0 for the other, is defined as the loss function. The forward model of the $\rm D^2NN$ is simulated based on angular spectrum propagation methods, and the errors are backpropagated to minimize the loss function via optimizing $\textbf{H}_i$, $a_i$, and $b_i$ by stochastic gradient descent algorithms.

\subsection*{Adaptive training process for the free-space DPU}
To deal with the systematic errors, in the experiments of 2-layer $\rm D^2NN$, we first deploy the network parameters of the first layer with the pre-trained model and capture the experimental outputs. Then the modulation coefficients of the second layer are re-trained with the first layer's experimental outputs to correct the first layer's systematic errors. To address the systematic errors of the second layer, for two intensities of the detection regions obtained from the output plane of the $\rm D^2NN$ after the experiments of the second layer, the intensity of first detector region was multiplied with a factor c within [0.9, 1.1]. $c$ was optimized to maximize the $F_{\beta}$ score on the training dataset, which was fixed in the inference process. This factor especially contributes to the errors caused by uneven light illumination.

\subsection*{Integrated DPU settings}
The integrated DPU comprises a 1-layer metaline at the working wavelength of 1550 nm for the on-chip epileptic seizure detection. The height of the meta-atom is fixed at 400 nm and the slot period is 300 nm. The slot width is chosen from 0 or 100 nm to implement binary modulation, and the corresponding phase modulation coefficients are 0 or -1.55 rad (see Supplementary Fig. 5 and Fig. 6). There are 1800 meta-atoms for processing EEG or iEEG signals, and DPU width is 540 $\rm\mu m$. The distance between the input plane and output plane is 200 $\rm \mu m$, and the metaline is placed at the center. The input waveguides are placed with an interval of 15 $\rm\mu m$, and the interval of two output waveguides is 270 $\rm\mu m$.

\subsection*{Training details}
All the simulation models were numerically implemented with Python (v3.6.13) and TensorFlow (v1.11.0) running on a desktop computer (Nvidia TITAN XP GPU, AMD Ryzen Threadripper 2990WX CPU with 32 cores, 128 GB of RAM, and the Microsoft Windows 10 operating system). The modulation coefficients of the diffractive layers were optimized via stochastic gradient descent and error backpropagation. The Adam optimizer \cite{kingma2014adam} and a learning rate of 0.01 were used with the loss functions of mean squared error or cross-entropy. Each $\rm D^2NN$ was trained with the epoch number of 1000. The training time of the free-space $\rm D^2NN$ was $\sim$4 h, and that for the integrated model was $\sim$1 h.


\backmatter

\bmhead{Data, materials, and code availability}
All data needed to evaluate the conclusions are present in the article and/or the Supplementary Information.

\bmhead{Acknowledgments}
This work is supported by the National Key Research and Development Program of China (No. 2021ZD0109902 and No. 2020AA0105500), the National Natural Science Foundation of China (No. 62088102 and No. 62275139), and the Tsinghua University Initiative Scientific Research Program.

\bmhead{Authors' contributions}
Q.D. and X.L. initiated and supervised the project. X.L. and T.Y. conceived the research. T.Y., X.L., M.Z., and X.C. designed the simulations and conducted the experiments. T.Y., X.L., M.Z., S.W., K.S., and H.Z. processed the data and analyzed the results. All authors prepared the manuscript and discussed the research.


\bibliography{sn-article}


\end{document}